# An Extended Symbol Table Infrastructure to Manage the Composition of Output-Specific Generator Information

Pedram Mir Seyed Nazari,[1] Alexander Roth,[1] Bernhard Rumpe[1]

**Abstract:** Code generation is regarded as an essential part of model-driven development (MDD) to systematically transform the abstract models to concrete code. One current challenges of template-based code generation is that output-specific information, i.e., information about the generated source code, is not explicitly modeled and, thus, not accessible during code generation. Existing approaches try to either parse the generated output or store it in a data structure before writing into a file. In this paper, we propose a first approach to explicitly model parts of the generated output. These modeled parts are stored in a symbol for efficient management. During code generation this information can be accessed to ensure that the composition of the overall generated source code is valid. We achieve this goal by creating a domain model of relevant generator output information, extending the symbol table to store this information, and adapt the overall code generation process.

## 1 Introduction

In model-driven development (MDD) code generation is an essential part to systematically generate detailed code from abstract input models. To bridge the gap between problem domain (abstract models) and solution domain (concrete code), MDD lifts the input models to primary artifacts in the development process. Regardless of the importance, code generator development is still a labor-intense and time-consuming task, where approaches to explicitly manage output-specific information are still lacking.

Explicitly management of output-specific code generator information, i.e., information about the generated source code, is essential for code generation to ensure that the generated source code is valid, i.e., well-formed. More importantly, it is necessary in the development process of code generators to split development tasks and in the maintenance phase as a documentation. For example, consider Java code is generated from UML class diagrams [Ru11]. In order to access parts of those generated Java classes are to be accessed during code generator runtime, the information of the relation of class diagram elements to Java elements is required such as class instantiation via the factory pattern [Ga95] versus direct instantiation via `new`-constructs.

Current code generator frameworks, e.g., [Me15, Xt15, Ac15, Je15], primarily focus on the code generation process and the development of code generators but mainly neglect explicit modeling of code generator output. Moreover, round-trip engineering [MER99] and reverse engineering [CC90] try to recreate models from the generated output. An approach to explicitly model information, which is exchanged during code generation, has been presented [JMS08]. However, the output still needs to either be first generated or approaches to address the required parts of the generated source code are lacking.

---

[1] RWTH Aachen University, Software Engineering, http://www.se-rwth.de

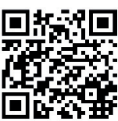

Hence, in this paper we present our approach to make output-specific code generator information explicit. As this information is dependent on the used input and output language, we present a preliminary domain model for a code generator that uses a variant of UML class diagrams [OM15] as input and Java as output. In this domain model, we use existing approaches [Ru11, Ru12] to map elements of UML class diagrams to Java code and provide an extension to manage Java object instantiation and field access via accessors and mutators. By making such information explicit, we enable code generator developers to exchange this information during development. Moreover, we make this information accessible at generation-time by extending the symbol table and the code generation process to allow storing arbitrary information.

Hence, we first introduce the basic concepts of symbol tables and code generation used in the MontiCore framework in Section 2. Then, we present our approach to manage output-specific code generator information by extending the symbol table and the code generation process (Section 3). Finally, we conclude our paper in Section 5.

## 2   Symbol Table and Code Generation

As the foundation for all aspects of language definition, language processing, and template-based code generation, we use the MontiCore language workbench [KRV10]. It uses a grammar defining the language to be processed and generates a parser and infrastructure for language processing based on this grammar. The generated infrastructure can be used to parse models conforming the defined grammar.

During processing of models the parser creates a *abstract syntax tree (AST)*, (an internal representation of the input model). This abstract representation is used for further phases of language processing, e.g., context condition checking and code generation. In addition, a *symbol table* is created in order to store relevant information for each model element.

### 2.1   Symbol Table

A symbol table (ST) is a data structure that maps names to associated model element information. In MontiCore, a *symbol* is an entry in the ST and represents a (named) model element [HNR15]. It contains all essential information related to that element. Different kinds of model elements, e.g., method and field in Java, are distinguished by corresponding *symbol kinds*. The main purpose of the symbol table is an efficient finding of model element specific information such as its type and its signature.

Compared to classical symbol tables, which are typically simple hash tables [Ah07], the symbol table in MontiCore is a combination of a (conceptual) table and the semantic model as described in [Fo10]. Its underlying infrastructure is a scope-tree containing a *collection* of symbols (cf. [Pa10]). Furthermore, it serves as a language-unspecific infrastructure for an efficient and effective integration of heterogeneous modeling languages [Lo13, Ha15])

Besides the information defined in the model element and represented by a symbol in the symbol table, a symbol can also contain information that is not explicitly part of that model element. For example, a Java field symbol can state whether it shadows a field of the super class or not. In addition, the source position of the model element can be stored in the symbol. Both information are not explicitly stated in the model element, but can be managed by the symbol. This allows to associate any kind of information—even technical information such as the source position— with the corresponding model element. We have even shown that a symbol table can manage code generator customizations [NRR15].

## 2.2    Code Generation

The code generation process of MontiCore is a mix of template-based and transformation-based code generation as shown in Figure 1. After the parser has created the AST, multiple transformations can be applied to transform the AST by adding, removing, or changing elements of the AST. The overall goal of the transformations is to make it fit the needed AST for code generation. During the transformation steps templates can be attached to AST elements in order to explicitly define the template to be used for this particular element. Certainly, this approach has limitations when generating non object-oriented code or when the input model is not a structural description that can be used for code generation. Thus, in the remainder of this paper we focus on a modeling language that is a variant of UML class diagrams [OM15] and Java as the output language of the code generator.

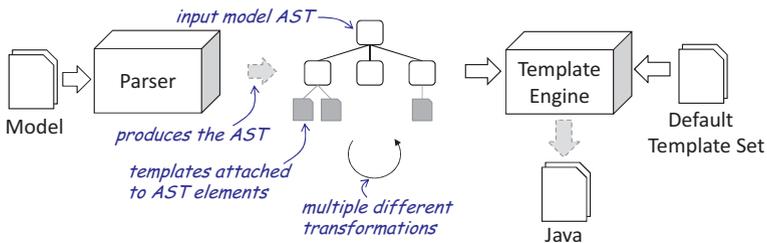

Fig. 1: Overview of a template-based and transformation-based code generation process.

After the transformations have been successfully applied, the AST is passed to the Template Engine. In addition, a default set of templates, which describe how to generate Java code from the input model, is passed to the template engine. When the template engine is started, it traverses the input AST and for each element executes either the attached templates of one of the default templates depending on the type of the AST element. Finally, the generated output is written to a file.

## 3    Managing Output-Specific Information with the Symbol Table

Our presented approach to manage output-specific code generator information is based on three elements. First, a common understanding of output-specific code generator information is needed. In general, this information is concerned with output language specific elements and concepts, e.g., object instantiation in Java. Second, an extension to the symbol table is required in order to add output-specific information and make it available at generation-time, i.e., run-time of the code generator. Third, the code generation approach needs to be adapted such that the information is added to the symbol table. Subsequently, we elaborate on each of the three main steps in more detail.

### 3.1   A Preliminary Domain Model for Class Diagrams and Java

A domain model of code generator output specific information depends on the output of the code generator and the input language. Hence, aiming for a general domain model for code generator output specific information is challenging and possibly not feasible. However, restricting the input language to UML class diagrams and the output language to Java, we try to provide a preliminary domain model that shows how code generator output-specific information can be modeled and managed with a symbol table. We do not claim for completeness of the domain model. Instead, we try to give an idea of how to model code generator output-specific information.

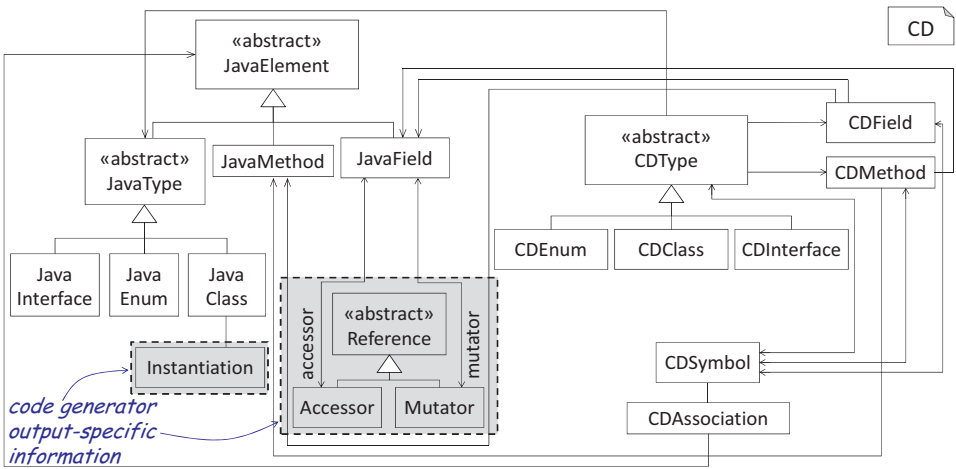

Fig. 2: Mapping of Symbols to Java Symbols and additional Generator Information

Our domain model in Figure 2 shows how UML class diagram symbols are mapped to Java symbols based on [Ru11, Ru12]. In this domain model, a `CDType`, which may represent a UML enum, interface or class, is mapped to a `JavaType`. We do not restrict the mapping to `JavaClasses`, because it may be necessary to generate interfaces or a modeled class. Moreover, each `CDField` is mapped to `JavaMethod` and `JavaField`. The mapping of a `CDField` to `JavaMethods` is optional as accessors and mutators may not be wanted. A `CDMethods` symbol is mapped to `JavaMethod` and `JavaField`. An example for a UML method that is mapped to a `JavaField` is an accessor that is mapped to the generated `JavaField` to allow for direct variable access.

Figure 2 gives an example for code generator output-specific information. This information is relevant for code generator developers and should be accessible during generation time rather than after code generation. Focusing on our small example, we have identified two types of code generator output-specific information. First, a `JavaField`, when mapped to Java code, can have `Accessors` and `Mutators`. This information is relevant during code generation as the generated code should access the field using the generated accessor and mutator. Thus, this information should be modeled explicitly and be accessible before the code is generated. Moreover, for `JavaClasses` the information relevant for creating instances of this class is required, e.g using the Singleton pattern [Ga95].

### 3.2 An Extension to the Symbol Table

Having an understanding of the mapping and additional information to be stored in the symbol table, we extend MontiCore's symbol table infrastructure to efficiently manage this information. The subsequent description is reduced to the essential parts and mainly focuses on the extensions, as shown in Figure 3. In [HNR15], we introduce the symbol table infrastructure in detail.

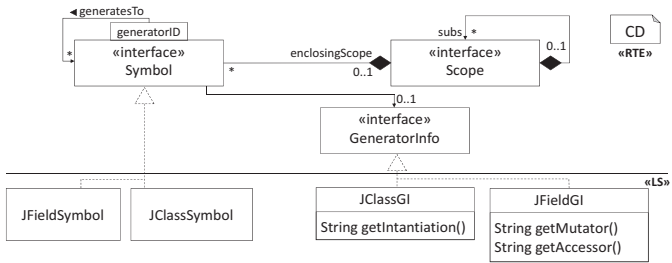

Fig. 3: Extended Symbol Table Infrastructure

As a first step, we enriched a symbol with information about the symbols of the target language it generates to. Moreover, symbols of the source language are associated with the corresponding symbols of the target language. For example, a class diagram field (source) can be generated as a Java field (target). Hence, the `CDFieldSymbol` maps to the corresponding `JavaFieldSymbol` (see Sect. 3). However, different generators can lead to different mappings and, thus, a unique generator id is used for each generator.

Second, each symbol now can optionally store generator-specific information, represented by the `GeneratorInfo` interface. `GeneratorInfo` must be implemented for each symbol of a target language and provide the required information. In the example of Java as the target language, information for, among others, classes and fields is needed, which are presented by `JavaClassGI` and `JavaFieldGI`, respectively. The former provides information such as how the generated class is instantiated, while the latter states how the field of the generated class is changed or accessed.

### 3.3 An Extension to Code Generation to handle Output-Specific Information

All modeled output-specific information is added to the symbol table to make it available at generation-time. Two different approaches can be used to add this information. First, before generation-time the transformations and templates can be parsed and the required information can be extracted. To identify relevant information comments or keywords may be used. While this approach makes sure that all information is available before generation-time, it has the disadvantage of processing all transformations and templates. In consequence, a supporting infrastructure to parse templates and transformations is necessary.

The second approach for adding all relevant information to the symbol table is to add it at generation-time. In particular, this means that all output-specific information is added to the symbol table while the code generation process is running. The information is not available before generation-time but still available at generation-time. A benefit of this approach is that no parsing of transformations and templates is required and the provided infrastructure can be kept small by providing an API to add this information. For our example in Figure 2, we can to provide the following methods:

- `toJavaType(CDType s, String className)`: Defines that a `CDType` is mapped to a `JavaType` with the name `className`.

- `toJavaField(CDField s, String fieldName)`: A mapping `CDField` to a `JavaField` is stored in the symbol table with the `fieldName` as the name of the generated Java field.

- `toJavaMethod(CDMethod s, String methodName)`: To define a mapping of a `CDMethod` to a `JavaMethod` the symbol table creates a Java method with the name `methodName`.

- `addInstantiation(JavaClass c, String code)`: In order to explicitly model object instantiation and store it in the symbol table, the API allows to add piece of `code` of type `String` to a Java class. For instance, to regard the Factory pattern, the piece of code can be "BookFactory.create()" for the Java class `Book`.

- `setAccessor(JavaField, String code)`: A mutator for a `JavaField` can be defined as a piece of `code` that represents, e.g., the name of the method ("getTitle" for a field named "title").

- `setMutator(JavaField, String code)`: For mutators the method is the same as for accessors. Additionally, we assume that each mutator requires one argument. Hence, when accessing this information in the symbol table a parameter should be passed. This is used to create the resulting string for the mutator.

A disadvantage is that the transformations and code templates have an execution order in which they have to be executed. If the execution order is violated, the information may not be available. In other words, the symbol in the symbol table cannot be resolved.

## 4   Related Work

Explicit modeling of output-specific code generator information is, to our knowledge, only hardly addressed by current literature. A closely related approach has been presented in [JMS08]. Here, a code generator is explicitly modeled via small interconnected services, which exchange information at runtime. This approach is similar to our presented approach as the exchanged information between serviced may contain generated informa-tion. In contrast, our presented approach proposes explicit modeling of this information and efficient management by using a symbol table.

Another approach that can be used to exchange information about the generate output has been presented by [ZR11]. The authors propose to generate the source code into containers before writing it into files. Hence, the complete source code is available at generation-time. However, as the authors are mainly concerned with producing syntactically correct output, there is no approach to address parts of the generated code as proposed by this paper. This is, however, essential to address composition of the generated source code, e.g., instantiation of generated Java classes.

Finally, an extension to round-trip engineering has been proposed to address the framework-provided abstractions via a dedicated domain-specific language (DSL) [AC06]. Rather than proposing a DSL, we explicitly model output-specific information using UML class diagrams and additionally provide efficient management at generation-time.

## 5  Conclusion

As code generation is regarded as an essential part of model-driven development to generate source code, output-specific code generator information has to be regarded in order to generate valid source code and decompose the generator development. In this paper, we presented a first approach to make output-specific code generator information explicit.

Our proposed approach consists of three steps. First, the relevant information is collected in a domain model. Based on this domain model the symbol table is extended to manage this information. Using the symbol table as an infrastructure has the benefit that the management is more efficient and no additional infrastructure is required. Finally, in the last step the code generation process needs to be adapted in order to make use of the stored information. We have applied this approach to a small use case to show how to model output-specific information for a UML class diagram to Java code generator. In particular, we focused on information related to object instantiation, and mutaturs and accessors for fields. In future, we plan to extend this approach to more real world examples.